\documentclass{article}
\usepackage{spconf,amsmath,graphicx,hyperref,amsfonts,amssymb, tabularx,fancyhdr}
\usepackage{footmisc}
\usepackage[misc]{ifsym}
\usepackage{lineno,hyperref}
\usepackage{multirow}
\usepackage{float}
\usepackage{booktabs}
\usepackage{xcolor}
\usepackage{url}
\usepackage{color}
\usepackage{array}
\usepackage{tikz}
\usepackage{colortbl}
\usepackage{mathrsfs,amsmath}
\usepackage{algorithm}
\usepackage{algorithmicx}
\usepackage{algpseudocode}
\usepackage{multirow}
\usepackage{float}
\usepackage{booktabs}
\fancypagestyle{firstpage}{
  \fancyhf{} 
  \fancyfoot[C]{\footnotesize © 20XX IEEE. Personal use of this material is permitted. 
  Permission from IEEE must be obtained for all other uses, in any current or future media, 
  including reprinting/republishing this material for advertising or promotional purposes, 
  creating new collective works, for resale or redistribution to servers or lists, or reuse 
  of any copyrighted component of this work in other works.}
}

\title{Uncertainty-Gated Deformable Network for Breast Tumor Segmentation in MR images}
\name{Yue Zhang$^{1\dag}$\textsuperscript{\Letter}\thanks{\dag Yue Zhang (yue\_zhang@uestc.edu.cn) and Jiahua Dong contribute equally to this work.}\quad Jiahua Dong$^{2\dag}$\quad Chengtao Peng$^{3}$\quad Qiuli Wang$^4$\quad Dan Song$^5$\quad Guiduo Duan$^{1}$}

\address{$^1$ National Key Laboratory of Intelligent Collaborative Computing,\\
University of Electronic Science and Technology of China, Chengdu, China\\
$^2$ College of Computer Science and Technology, Zhejiang University, Hangzhou, China\\
$^3$ School of Computer Science and Information Engineering, Hefei University of Science, Hefei, China
\\
$^4$ Department of Radiology, Southwest Hospital, Army Medical University, Chongqing, China\\
$^5$ School of Electrical and Information Engineering, Tianjin University, Tianjin, China}

\begin{document}
\ninept
\maketitle
\thispagestyle{firstpage}
\begin{abstract}
Accurate segmentation of breast tumors in magnetic resonance images (MRI) is essential for breast cancer diagnosis, yet existing methods face challenges in capturing irregular tumor shapes and effectively integrating local and global features. To address these limitations, we propose an uncertainty-gated deformable network to leverage the complementary information from CNN and Transformers. Specifically, we incorporates deformable feature modeling into both convolution and attention modules, enabling adaptive receptive fields for irregular tumor contours. We also design an Uncertainty-Gated Enhancing Module (U-GEM) to selectively exchange complementary features between CNN and Transformer based on pixel-wise uncertainty, enhancing both local and global representations. Additionally, a Boundary-sensitive Deep Supervision Loss is introduced to further improve tumor boundary delineation. Comprehensive experiments on two clinical breast MRI datasets demonstrate that our method achieves superior segmentation performance compared with state-of-the-art methods, highlighting its clinical potential for accurate breast tumor delineation.
\end{abstract}
\section{Introduction}
\label{sec:intro}
Breast cancer is a leading cause of cancer-related mortality among women. Among imaging modalities, magnetic resonance imaging (MRI) is particularly valuable due to its high soft-tissue contrast, which facilitate reliable detection and evaluation of breast tumors. Accurate tumor segmentation from breast MRI is essential in clinical practice, as it enables lesion localization and supports quantitative analysis.
Traditional manual delineation of tumor boundaries, however, is time-consuming and subject to inter-observer variability. This underscores the need for automated and accurate segmentation methods for breast tumors.

In recent years, deep learning has greatly advanced breast tumor segmentation. Convolutional neural networks (CNNs)~\cite{ronneberger2015u,zhou2018unet++,huang2020unet,isensee2021nnu,guo2022unet,chen2022aau,wang2024progressive,peng2022imiin} excel at capturing local structural details but are limited by their inductive biases. Transformer-based networks~\cite{cao2022swin,strudel2021segmenter,huang2021missformer,wu2023d} effectively model global context but often fail to preserve fine-grained local information. To address these complementary limitations, hybrid CNN–Transformer networks~\cite{chen2021transunet,zheng2021rethinking,hatamizadeh2022unetr,zhou2023nnformer,liu2022phtrans,he2023h2former,li2025cfformer,yuan2023effective} have been developed to leverage the strengths of both paradigms.
Typical hybrid designs include \emph{using a Transformer encoder with a CNN decoder}~\cite{chen2021transunet}, \emph{integrating Transformers into high-level layers}~\cite{yuan2023effective}, or \emph{embedding CNN and Transformer modules jointly at multiple stages}~\cite{li2025cfformer} either in a sequential or parallel manner.

Despite their advantages, existing hybrid models still face several limitations. First, the interaction between CNN and Transformer features remains suboptimal, leading to redundant information exchange. It is intuitive that not all regions in the MRI require global context, and part of regions rely primarily on local information that CNNs can capture efficiently. Conversely, Transformer features do not always need fine-grained local details. This highlights the need for a selective mechanism to adaptively balance local and global representations. Second, existing models often overlook the irregular shapes and wide size variations of breast tumors. Fixed rectangular receptive fields in conventional convolution and attention mechanisms are insufficient to flexibly capture such shape characteristics.

\begin{figure*}[t]
\centering
\includegraphics[width=0.95\textwidth]{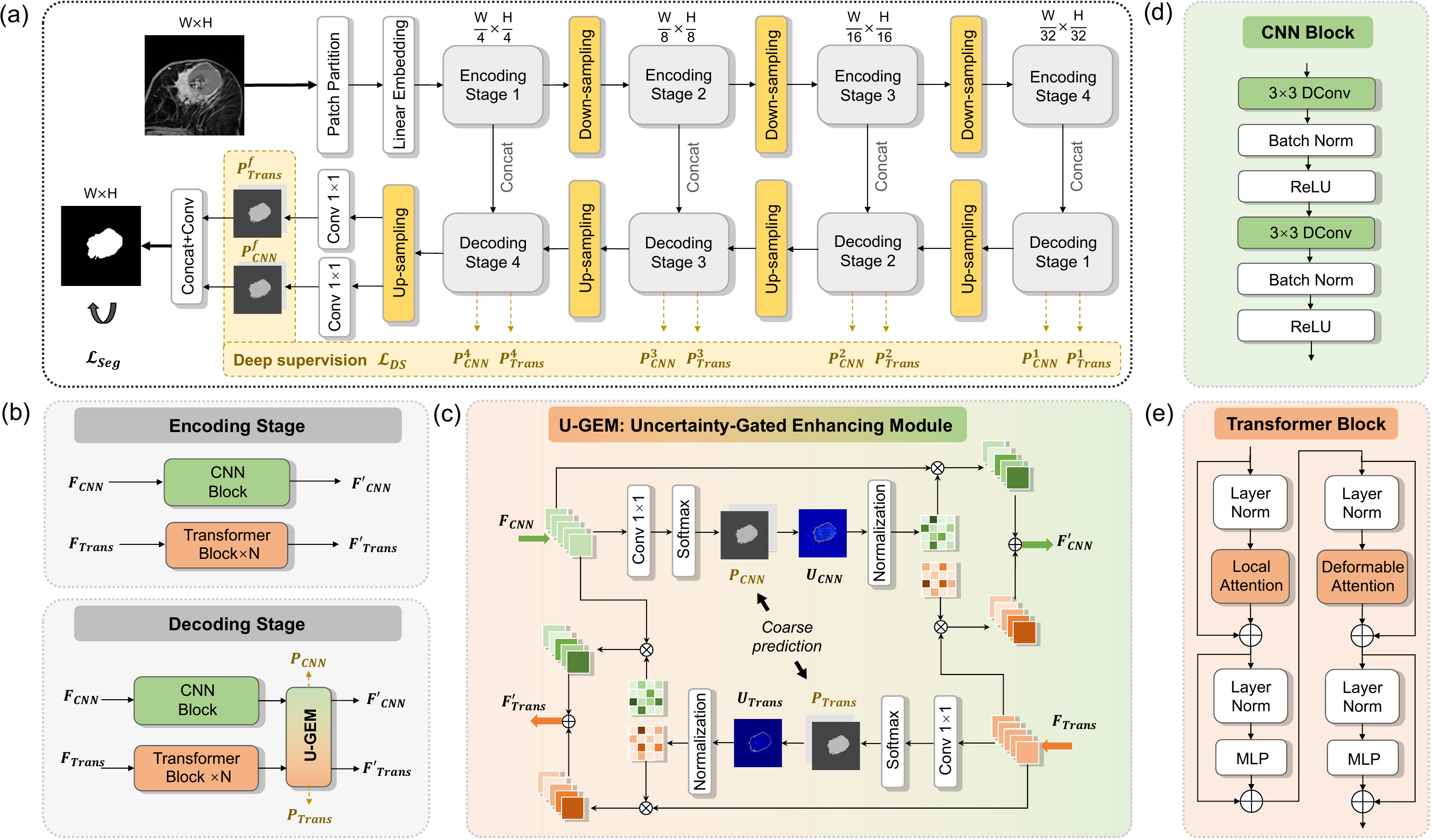}
\caption{Schematic illustration of the proposed network for breast tumor segmentation in MR images. The model enhances both CNN and Transformer branches while adapting to irregular tumor shapes. (a) Overall network architecture based on an encoder–decoder framework. (b) Structure of the encoding and decoding stages. (c) Design of U-GEM, which enables information exchange and mutual enhancement between the CNN and Transformer branches. (d) Details of the CNN block. (e) Details of the Transformer block.
} \label{fig_overview}
\end{figure*}

To address these limitations, we propose an uncertainty-gated deformable network within a hybrid CNN-Transformer architecture. First, we introduce deformable feature modeling into both convolutional and attention modules, enabling adaptive receptive field that better captures irregular tumor contours.
Second, an Uncertainty-Gated Enhancing Module (U-GEM) is designed to exchange information between CNN and Transformer branches. U-GEM uses the entropy-based uncertainty as gating signals, and selectively incorporates complementary features from another branch only in regions of high uncertainty, thereby achieving mutual enhancement of both branches.
Apart from this, a Boundary-sensitive Deep Supervision Loss (BDS-Loss) is specially designed to fully optimize the hybrid network and improve tumor boundary delineation.
Comprehensive experiments on two clinical breast MR datasets demonstrates the superiority of our method over state-of-the-art methods.
The main contributions of this work are summarized as follows:
\begin{itemize}
    \item We incorporate deformable feature modeling into CNN and Transformer blocks, which enhances the ability to capture tumors with irregular boundaries and large shape variations.
    \item We propose the U-GEM for hybrid CNN-Transformer architecture, which allows adaptive and uncertainty-gated feature enhancement between local and global representations.
    \item The proposed method achieves superior segmentation performance on two MR datasets, highlighting its clinical potential for accurate breast tumor delineation.
\end{itemize}

\section{Method}
\label{method}

\subsection{Model Architectures}
In this work, we propose an uncertainty-gated deformable network for breast tumor segmentation in MR images. As illustrated in Fig.~\ref{fig_overview}(a), the network follows an encoder–decoder design, where the encoder consists of four hierarchical encoding stages. Each encoding stage consists a CNN block (see Fig.~\ref{fig_overview}(d)) and a Transformer block (see Fig.~\ref{fig_overview}(e)) to capture both local texture details and long-range dependencies. Specifically, deformable operations are introduced in CNN/Transformer blocks to flexibly model irregular tumor shapes. In the decoder, four corresponding decoding stages are employed, where CNN and Transformer features are mutually enhanced via four Uncertainty-Gated Enhancing Modules (U-GEMs). U-GEM enables mutual guidance between the two feature domains, enhancing robustness and improving feature extraction. Finally, the network outputs segmentation maps from both CNN and Transformer branches, as well as a fused final prediction, ensuring accurate and reliable tumor segmentation.

\subsection{Deformable Feature Modeling}
Considering the irregular shapes and large size variations of breast tumors in MRI, we introduce the mechanism of deformable receptive field into the design of both CNN and Transformer blocks. This ensures that our model can flexibly capture tumor contours.


As shown in Fig.~\ref{fig_overview}(d), our CNN block consists of two consecutive groups of DConv (deformable convolution~\cite{dai2017deformable})-BN-ReLU layers. In contrast to traditional convolution that samples at fixed offsets, deformable convolution augments these locations with learnable offsets, which enables the kernel to adapt to semantically aligned regions with irregular shapes.
Besides, the Transformer block consists of successive local attention~\cite{hassani2023neighborhood} and deformable attention~\cite{xia2022vision} blocks (see Fig.~\ref{fig_overview}(e)).
Similarly, deformable attention samples keys and values from a few content-adaptive reference points rather than fixed window regions, enabling the Transformer to flexibly focuses attention on informative regions.


\subsection{Uncertainty-Gated Enhancing Module}
To effectively integrate the local modeling capability of CNNs and the long-range dependency modeling of Transformers, we introduce a U-GEM between CNN and Transformer blocks. This module adaptively guides information exchange by estimating and leveraging the pixel-wise uncertainty of each branch. In this way, both CNN and Transformer branches are supplemented and enhanced.

\begin{figure*}[t]
\centering
\includegraphics[width=0.98\textwidth]{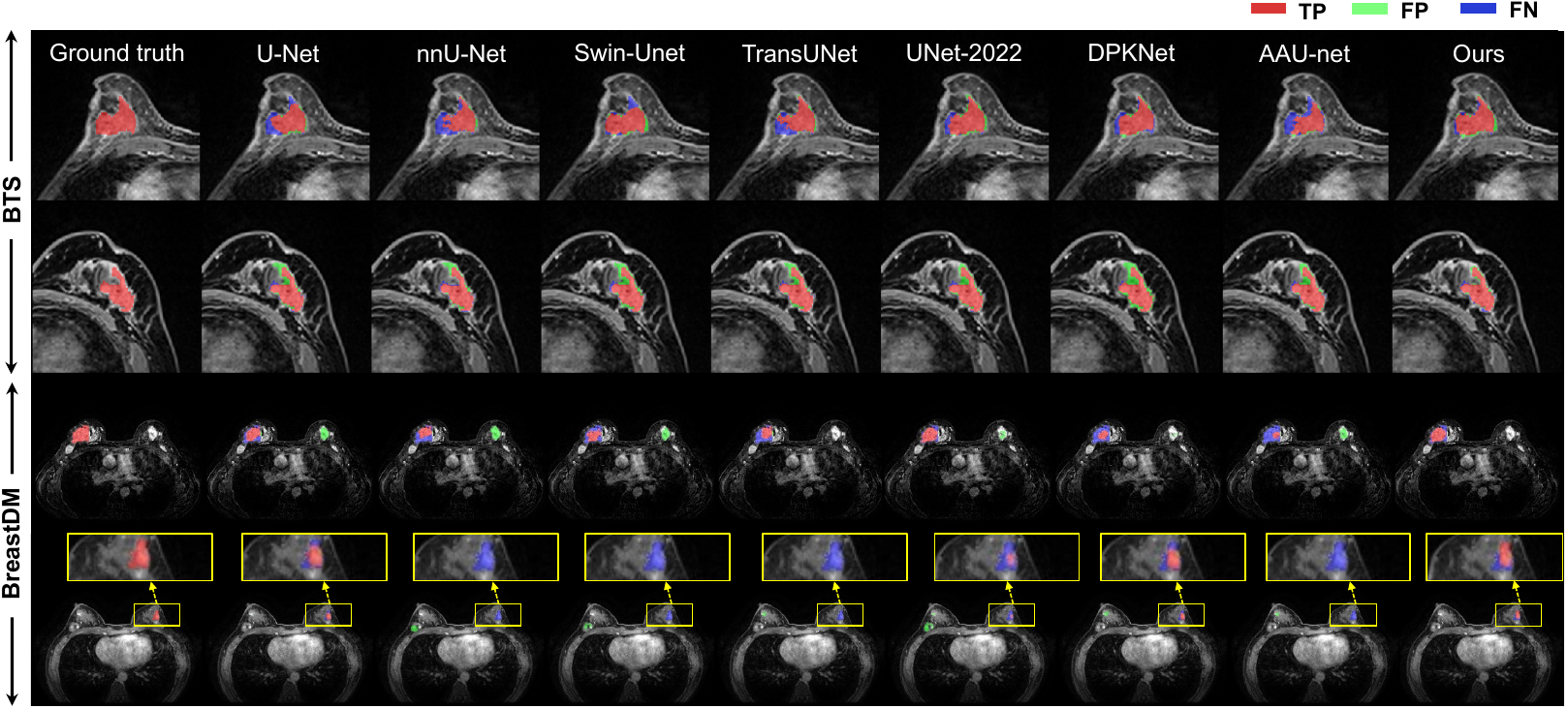}
\caption{Qualitative results on the BTS and BreastDM datasets. The \textcolor{red}{red}, \textcolor{green}{green}, and \textcolor{blue}{blue} regions indicate true positives (TP), false positives (FP), and false negatives (FN), respectively. The \textcolor{yellow}{yellow} boxed region is enlarged to facilitate a clearer comparison for small breast tumors.
} \label{fig_results}
\end{figure*}

\textbf{Transformer Assisting CNN:}
First, uncertain regions in the CNN features request complementary information from the Transformer features.
Let \( F_{\text{CNN}} \in \mathbb{R}^{C \times H \times W} \) and \( F_{\text{Trans}} \in \mathbb{R}^{C \times H \times W} \) denote the feature maps from the CNN and Transformer branches. 
We compute a coarse prediction $P_{\text{CNN}}$ from the CNN features via a classifier head:
$
P_{\text{CNN}} = \text{Softmax}( \text{Conv}_{1 \times 1}(F_{\text{CNN}}) ).
$
Then, an uncertainty map \( U_{\text{CNN}} \) is computed using pixel-wise entropy:
\[
U_{\text{CNN}}(i,j) = \Big(- \sum_{c=1}^{C} P_{\text{CNN}}^{(c)}(i,j) \log P_{\text{CNN}}^{(c)}(i,j)\Big)\cdot(1/\log|C|),
\]
where $C$ denotes the number of classes. Pixels with higher uncertainty values indicate greater classification ambiguity, which requires additional guidance from the Transformer branch to support the decision. Based on $U_{CNN}$, the enhanced CNN feature is computed as:
\[
F'_{\text{CNN}} = (1 - U_{\text{CNN}}) \cdot F_{\text{CNN}} + U_{\text{CNN}} \cdot F_{\text{Trans}},
\]

\textbf{CNN Assisting Transformer:}
In turn, uncertain regions in the Transformer features request support from the CNN features.
Similarly, the Transformer branch generates its coarse prediction $P_{\text{Trans}}$ following the same procedure, and the corresponding uncertainty map $U_{\text{Trans}}$ is then computed.
Then, the feature enhancing in the opposite direction is:
\[
F'_{\text{Trans}} = (1 - U_{\text{Trans}}) \cdot F_{\text{Trans}} + U_{\text{Trans}} \cdot F_{\text{CNN}},
\]
The bidirectional enhanced outputs \( F'_{\text{CNN}} \) and \( F'_{\text{Trans}} \) are forwarded into their respective branches. This design allows each branch to selectively absorb complementary information where its own predictions are uncertain: enhancing CNNs in modeling global context and improving Transformers in capturing local details.

\subsection{Boundary-Sensitive Deep Supervision Loss}
To effectively guide network optimization, we design a composite loss termed \emph{BDS-Loss} (Boundary-sensitive Deep Supervision Loss). 
It is composed of two parts: a deep supervision loss $\mathcal{L}_{DS}$ and a segmentation loss 
$\mathcal{L}_{Seg}$. 
The deep supervision strategy is applied to guide the intermediate predictions generated by 
U-GEMs, together with the final outputs of CNN and Transformer branches. 
All these outputs are supervised by a boundary-sensitive BoundaryDoU~\cite{sun2023boundary} loss (BDoU):
\begin{align*}
\mathcal{L}_{DS} = \sum_{i=1}^{4} \Big( w_{c}^i\cdot\text{BDoU}(P^{i}_{CNN}, Y) + w_{t}^i\cdot\text{BDoU}(P^{i}_{Trans}, Y) \Big) \\
+ w_{c}^f\cdot\text{BDoU}(P^{f}_{CNN}, Y) + w_{t}^f\cdot\text{BDoU}(P^{f}_{Trans}, Y),
\end{align*}
where $P^{i}_{CNN}$ and $P^{i}_{Trans}$ denote the $i$-th intermediate predictions, and $P^{f}_{CNN}$ and $P^{f}_{Trans}$ correspond to the final outputs of the CNN and Transformer branches, respectively. 
The weighting coefficients $\{w_c^1,w_c^2,w_c^3,w_c^4,w_c^f\}$ and $\{w_t^1,w_t^2,w_t^3,w_t^4,w_t^f\}$ are used to balance the contributions of different predictions, both configured as $\{0.1,0.2,0.5,0.8,1\}$.

\begin{table*}
\centering
\small
\caption{Quantitative results on the BTS dataset and BreastDM dataset.}
\label{tb_results}
\begin{tabular}{ccccccc|cccccc} 
\hline
\multicolumn{7}{c|}{\textbf{BTS dataset}} & \multicolumn{6}{c}{\textbf{BreastDM dataset}} \\ 
\hline
\textbf{Method} & \begin{tabular}[c]{@{}c@{}}\textbf{DSC$\uparrow$}\\\textbf{(\%)}\end{tabular} & \begin{tabular}[c]{@{}c@{}}\textbf{HD$\downarrow$}\\\textbf{(mm)}\end{tabular} & \begin{tabular}[c]{@{}c@{}}\textbf{ASD$\downarrow$}\\\textbf{(mm)}\end{tabular} & \begin{tabular}[c]{@{}c@{}}\textbf{ASSD$\downarrow$}\\\textbf{(mm)}\end{tabular} & \begin{tabular}[c]{@{}c@{}}\textbf{Sens.$\uparrow$}\\\textbf{(\%)}\end{tabular} & \begin{tabular}[c]{@{}c@{}}\textbf{Prec.$\uparrow$}\\\textbf{(\%)}\end{tabular} & \begin{tabular}[c]{@{}c@{}}\textbf{DSC$\uparrow$}\\\textbf{(\%)}\end{tabular} & \begin{tabular}[c]{@{}c@{}}\textbf{HD$\downarrow$}\\\textbf{(mm)}\end{tabular} & \begin{tabular}[c]{@{}c@{}}\textbf{ASD$\downarrow$}\\\textbf{(mm)}\end{tabular} & \begin{tabular}[c]{@{}c@{}}\textbf{ASSD$\downarrow$}\\\textbf{(mm)}\end{tabular} & \begin{tabular}[c]{@{}c@{}}\textbf{Sens.$\uparrow$}\\\textbf{(\%)}\end{tabular} & \begin{tabular}[c]{@{}c@{}}\textbf{Prec.$\uparrow$}\\\textbf{(\%)}\end{tabular} \\ 
\hline
U-Net~\cite{ronneberger2015u} & 76.44 & 9.47~ & 2.72 & 1.65~ & 80.36 & 77.48 & 64.53 & 28.67~ & 26.61 & 26.60~ & 60.33 & 77.13 \\
nnU-Net~\cite{isensee2021nnu} & 79.83 & 4.25~ & 1.15 & 0.84~ & 83.10 & 80.43 & 70.02 & 25.81~ & 23.91 & 24.04~ & 67.59 & 79.43 \\
Swin-Unet~\cite{cao2022swin} & 79.10 & 3.59~ & 0.92 & 0.76~ & 85.10 & 77.19 & 70.41 & 17.73~ & 15.51 & 15.49~ & 66.29 & 82.52 \\
TransUNet~\cite{chen2021transunet} & 81.44 & 3.42~ & 0.80 & 0.73~ & 84.64 & \textbf{81.61} & 69.08 & 18.68~ & 16.50 & 16.70~ & 65.72 & 82.40 \\
UNet-2022~\cite{guo2022unet} & 80.36 & 3.48~ & 0.80 & 0.71~ & 85.75 & 78.60 & 73.57 & 16.49~ & 14.78 & 14.75~ & \textbf{73.08} & 79.40 \\
DPKNet~\cite{wang2024progressive} & 74.72 & 6.07~ & 1.90 & 1.22~ & 92.20 & 65.74 & 71.96 & 16.51~ & 14.44 & 14.48~ & 68.71 & 82.05 \\
AAU-net~\cite{chen2022aau} & 76.93 & 5.06~ & 1.42 & 1.02~ & 80.68 & 77.84 & 73.32 & 9.13~ & 6.97 & 7.02~ & 69.95 & \textbf{84.17} \\
\textbf{Ours} & \textbf{82.63} & \textbf{2.87~} & \textbf{0.67} & \textbf{0.66~} & \textbf{87.46} & 81.02 & \textbf{74.65} & \textbf{7.41~} & \textbf{5.37} & \textbf{5.30~} & 72.68 & 82.95 \\
\hline
\end{tabular}
\end{table*}
\begin{table*}[t]
\centering
\small
\caption{Ablation study on the BTS dataset. `H': a simple CNN-Transformer hybrid architecture. `D': deformable feature modeling. `U': U-GEM. `B': BDS-Loss.}
\label{tb_ablation}
\begin{tabular}{lcccccc} 
\hline
 & \textbf{DSC $\uparrow$(\%)} & \textbf{HD $\downarrow$(mm)} & \textbf{ASD $\downarrow$(mm)} & \textbf{ASSD $\downarrow$(mm)} & \textbf{Sens. $\uparrow$(\%)} & \textbf{Prec. $\uparrow$(\%)} \\ 
\hline
Baseline (Swin-Unet) & 79.10 & 3.59~ & 0.92~ & 0.76~ & 85.10 & 77.19 \\
H & 80.28 & 3.39~ & 0.76~ & 0.76~ & 80.24 & \textbf{83.65} \\
H+D & 81.61 & 3.10~ & 0.71~ & 0.69~ & 86.06 & 80.72 \\
H+D+U &  82.11&  3.04~ &  0.70~ &  0.69~ &  86.85&  80.59\\
\textbf{H+D+U+B} & \textbf{82.63} & \textbf{2.87~} &  \textbf{0.67~} & \textbf{0.66~} & \textbf{87.46} & 81.02 \\
\hline
\end{tabular}
\end{table*}

The segmentation loss $\mathcal{L}_{Seg}$ is applied to the final fused output of the network to ensure accurate delineation of tumor boundaries.
It integrates two objectives: the Dice loss, and the BoundaryDoULoss\cite{sun2023boundary} that explicitly optimizes boundary localization:
\[
\mathcal{L}_{Seg} = \text{Dice}(P, Y)+\text{BDoU}(P, Y).
\]
where $P$ is the final prediction and $Y$ is the ground truth.

\section{Experiments}
\label{sec:pagestyle}
\subsection{Experimental Setup}
To comprehensively validate the proposed method, we conduct extensive experiments on two breast MR datasets: the BTS~\cite{peng2022imiin} dataset and BreastDM~\cite{zhao2023breastdm} dataset. The BTS dataset is a clinical 2D MRI breast tumor dataset collected in our previous work~\cite{peng2022imiin}, which includes 432 T1C cases. The BreastDM dataset is a publicly available DCE-MRI breast tumor dataset that contains 235 cases. For both datasets, all images are resized to a size of 224×224. All samples are split into training, validation, and testing sets with a ratio of 7:1:2.

For comparison, we select seven representative methods, including five general segmentation networks: U-Net~\cite{ronneberger2015u}, nnU-Net~\cite{isensee2021nnu}, Swin-Unet~\cite{cao2022swin}, TransUNet~\cite{chen2021transunet}, and UNet-2022~\cite{guo2022unet}, as well as two state-of-the-art breast tumor segmentation networks: DPKNet~\cite{wang2024progressive} and AAU-net~\cite{chen2022aau}. All networks are trained for 150 epochs on a single NVIDIA RTX 3090 GPU. The training is performed using the SGD optimizer with a momentum of 0.9 and a weight decay of 0.0001. The batch size is set to 16, and the initial learning rate is 0.001, which is gradually decayed to zero.

Six common segmentation metrics are used for evaluation: The Dice Similarity Coefficient (DSC) that measures region overlap; the Hausdorff Distance (HD), Average Surface Distance (ASD) and Average Symmetric Surface Distance (ASSD) that quantify boundary distances; Sensitivity (Sens.) that evaluates tumor detection, and Precision (Prec.) that evaluates the pixel-wise classification.

\subsection{Results}
As reported in Table~\ref{tb_results}, the proposed method achieves the highest scores on most metrics across both datasets. On the BTS dataset, our model reaches a DSC of 82.63\%, surpassing the second-best method (TransUNet) by 1.19\%. Coupled with the lowest HD (2.87 mm), ASD (0.67 mm) and ASSD (0.66 mm), it is indicated that our method is able to accurately delineate tumors with irregular and subtle contours. On the BreastDM dataset, our model yields a DSC of 74.65\%, outperforming the strongest competing method (AAU-net) by 1.33\%. Our method also achieves the lowest boundary distances. Considering the inherent challenges of BreastDM, including lower contrast and smaller tumor size, this improvement demonstrates the robustness of our hybrid CNN–Transformer architecture.

Fig.~\ref{fig_results} further illustrates the visual comparison results on two datasets. On the BTS dataset, competing methods often exhibit false negatives at tumor boundaries, where subtle enhancing regions are omitted, or false positives in surrounding glandular tissues (also reflected in Sens. and Prec. metrics). In contrast, our model better preserves complete tumor structures, which reflects the effectiveness of deformable feature modeling and uncertainty-gated feature enhancing. On the BreastDM dataset, the challenges are more pronounced: smaller tumors and heterogeneous backgrounds increase the likelihood of missed detections. Nevertheless, our method demonstrates more faithful localization with fewer false positives/negatives. Particularly, in the forth row of Fig.~\ref{fig_results}, showing a small tumor case from the BreastDM, most competing methods fail to detect the lesion or produce false positives. By contrast, our method successfully delineates the small tumor with accurate shape, demonstrating its sensitivity to subtle lesions.

\subsection{Ablation Study}
To verify each component within our network, we start from a pure Transformer network, i.e., the \textbf{Baseline} (Swin-Unet), which achieves a DSC of 79.10\%. We gradually add components to the Baseline and summarize the results in Table~\ref{tb_ablation}.

\textbf{Effectiveness of CNN features:}
By incorporating a CNN branch (using vanilla convolution) to the Transformer baseline, we obtain a simple hybrid network (\textbf{H}). The CNN branch complements the Transformer with stronger local feature extraction, and the DSC increases to 80.28\% (+1.18\%).

\textbf{Effectiveness of deformable feature modeling:}
Introducing the deformable feature modeling (\textbf{D}) raises the DSC to 81.61\% (+1.33\%). Deformable operations yield clear improvements as these operators are capable of capturing irregular tumor shapes.

\textbf{Effectiveness of U-GEM:}
By incorporating the U-GEMs (\textbf{U}), the DSC further increases to 82.11\% (+0.50\%). This gain demonstrates the effectiveness of uncertainty-gated feature exchange between CNN and Transformer pathways. Uncertain regions in one branch are compensated by the other, hence improving robustness in challenging cases.

\textbf{Effectiveness of BDS-Loss:}
Finally, adopting the proposed BDS-loss (\textbf{B}) achieves the best DSC of 82.63\% (+0.52\%). Three boundary metrics are further reduced, which confirms the effectiveness of explicit boundary supervision.


\section{Conclusion}
In this work, we presented an uncertainty-gated deformable network for breast tumor segmentation in MR images. Building upon a hybrid CNN-Transformer architecture, we integrated deformable feature modeling, the Uncertainty-Gated Enhancing Module, and a BDS-Loss. Our method effectively captures both local details and global context while adapting to irregular tumor shapes. Experiments on two datasets demonstrated consistent improvements in common metrics, which confirms the effectiveness and superiority of our method for breast tumor segmentation.

\bibliographystyle{IEEEbib}
\bibliography{refs}

\end{document}